\documentclass[aps,prd,preprintnumbers,superscriptaddress,nofootinbib]{revtex4}

\usepackage{hyperref}
\hypersetup{colorlinks=true,linkcolor=purple,anchorcolor=blue,citecolor=blue, filecolor=blue,urlcolor=red,bookmarksnumbered=true,
	pdfview=FitB
}
\usepackage{color}
\usepackage{xcolor}
\usepackage{ulem}
\usepackage{csquotes}
\colorlet{purple1}{blue!70!red}
\colorlet{darkred}{red!50!black}
\usepackage{graphicx}
\usepackage{psfrag}
\usepackage{color}
\usepackage{comment}
\usepackage{amssymb}
\usepackage{amsmath,amssymb}
\usepackage{mathtools}
\usepackage{float}
\usepackage{color}
\usepackage{leftidx}
\usepackage{booktabs}
\usepackage{latexsym}
\usepackage{revsymb}
\usepackage{multirow}
\usepackage{hypcap}
\usepackage{array}
\usepackage[english]{babel}
\usepackage{amsmath}
\usepackage{epstopdf}
\usepackage{natbib}
\usepackage{comment}

\begin{document}
	
\title{3D EMT distributions as an Abel image of 2D EMT distributions on the light front}
\author{Poonam Choudhary}
\email{poonamch@iitk.ac.in}
\author{Bheemsehan~Gurjar}
\email{gbheem@iitk.ac.in} 
\author{Dipankar~Chakrabarti}
\email{dipankar@iitk.ac.in} 
\affiliation{Indian Institute of Technology Kanpur, Kanpur-208016, India}
\author{Asmita Mukherjee}
\email{asmita@phy.iitb.ac.in}
\affiliation{\small{Department of Physics, Indian Institute of Technology Bombay,Powai, Mumbai 400076, India}}

\date{\today}

	
	\begin{abstract}
		The energy-momentum tensor (EMT) and corresponding gravitational form factors (GFFs) provide
us information about the internal structure like spin, mass and spatial densities of the proton. The
Druck gravitational (D-term) form factor is related to the mechanical stability of the proton and gives
information about the spatial distributions of the forces inside the hadron. In this work, we study the
GFFs in the framework of the light-front quark diquark model. The model has been successful to derive
various properties of protons. We investigate the three-dimensional spatial distributions of proton as
an Abel image of two-dimensional distributions in this model[1]. We explicitly show the global and local
stability conditions which are satisfied by both 2D and 3D distributions in our model. We compare our
results with the chiral quark soliton model, JLab and lattice data.\\

\noindent \bf{\textit{Presented at DIS2022: XXIX International Workshop on Deep-Inelastic Scattering and Related Subjects, Santiago de Compostela, Spain, May 2-6 2022. }}
	\end{abstract}
\maketitle	
	\section{Introduction}
	\label{sec:Introduction}
 The scattering of the proton by gravitational field is described by gravitational form factors (GFFs) which explains mass, spin and force distributions inside the proton~\cite{Harindranath:2013goa,Pagels:1966zza}. Gravitational form factors are parameterized in terms of the matrix element of the energy-momentum tensor between the incoming and outgoing proton states. Each element of the energy-momentum tensor give information about matter coupling to the gravitational field. The total symmetric EMT for a system (quarks and gluons) can be parameterized in terms of three GFFs: $A(q^2)$, $J(q^2)=(1/2)(A(q^2)+B(q^2))$ and $D(q^2)$ as~\cite{Ji:2012vj}
	 \begin{eqnarray}\label{QCDEMT}
		\left\langle p^{\prime}\left|\hat{\Theta}_{\mathrm{QCD}}^{\mu \nu}(0)\right| p\right\rangle=& \bar{u}\left(p^{\prime}\right)\left[A(q^2) \frac{P^{\mu} P^{\nu}}{M}+J(q^2) \frac{i P^{\{\mu} \sigma^{\nu\}{\alpha}}  \Delta_{\alpha}}{M}\right.\nonumber \\
		&\left.+\frac{D(q^2)}{4 M}\left(\Delta^{\mu} \Delta^{\nu}-\eta^{\mu \nu} \Delta^{2}\right)\right] u(p),
	\end{eqnarray}	
	
 The GFFs contains the essential information on the internal
structure of the proton and could be extracted through hard exclusive processes like deeply virtual Compton scattering as the second moments of Generalized Parton distribution functions (GPDs)~\cite{Burkert:2018bqq,Polyakov:2002yz}. The GFFs $A(q^{2})$ and $J(q^{2})$ give the mass and angular momentum of the proton and are constrained at $q^{2}=0$, i.e., $A(0)=1$ and $J(0)=(1/2)(A(0)+B(0))=1/2$~\cite{Ji:1996ek}. While, The $ \textit{D-term} $, which is related to the mechanical properties  of the proton, is extracted through the spatial-spatial component of the energy-momentum tensor, is deeply related to the stability of the proton and is unconstrained at $q^{2} = 0$~\cite{Polyakov:2018zvc,Lorce:2018egm}.

	\section{Light front quark diquark model}\label{LFQDQmodel}
In quark-diquark model, We assume that the virtual incoming photon is interacting with a active valence quark and the two other spectator valence quarks form a diquark of spin 0, called a scalar diquark. Therefore the proton state $|P,S\rangle$ having momentum $P$ and spin $S$, can be represented as a two-particle Fock-state as following
	\begin{eqnarray} \label{protonstate}
	|P ; \pm\rangle&=& \sum_{q} \int \frac{d x d^{2} \mathbf{p}_{\perp}}{2(2 \pi)^{3} \sqrt{x(1-x)}}\nonumber \\  
	&&\times\left[\psi_{+}^{q \pm}\left(x, \mathbf{p}_{\perp}\right)\left|+\frac{1}{2}, 0 ; x P^{+}, \mathbf{p}_{\perp}\right\rangle+\psi_{-}^{q \pm}\left(x, \mathbf{p}_{\perp}\right)\left|-\frac{1}{2}, 0 ; x P^{+}, \mathbf{p}_{\perp}\right\rangle\right], 
	\end{eqnarray}
	$\psi_{\lambda_{q}}^{q \lambda_{N}}$ are light-front wave functions which are given ~\cite{Gutsche:2013zia} by following expressions:
	\begin{eqnarray}\label{LFWF}
	\psi_{+}^{q+}\left(x, \mathbf{p}_{\perp}\right)&=& \varphi^{q(1)}\left(x, \mathbf{p}_{\perp}\right) ,\hspace{1cm}
	 \psi_{-}^{q+}\left(x, \mathbf{p}_{\perp}\right)=-\frac{p^{1}+i p^{2}}{x M} \varphi^{q(2)}\left(x, \mathbf{p}_{\perp}\right) \nonumber \\ 
	\psi_{+}^{q-}\left(x, \mathbf{p}_{\perp}\right)&=& \frac{p^{1}-i p^{2}}{x M} \varphi^{q(2)}\left(x, \mathbf{p}_{\perp}\right), \hspace{1cm}
	\psi_{-}^{q-}\left(x, \mathbf{p}_{\perp}\right)= \varphi^{q(1)}\left(x, \mathbf{p}_{\perp}\right) 
	\end{eqnarray}
	
	where $\varphi_{q}^{(i=1,2)}(x,\mathbf{p}_{\perp})$ are the wave functions predicted by the soft-wall AdS/QCD and can be written as~\cite{Brodsky:2006ha}
	\begin{eqnarray}  \label{phiAdSQCD}
	\varphi^{q(i)}\left(x, \mathbf{p}_{\perp}\right)=N_{q}^{(i)} \frac{4 \pi}{\kappa} \sqrt{\frac{\log (1 / x)}{1-x}} x^{a_{q}^{(i)}}(1-x)^{b_{q}^{(i)}} \exp \left[-\frac{\mathbf{p}_{\perp}^{2}}{2 \kappa^{2}} \frac{\log (1 / x)}{(1-x)^{2}}\right];
	\end{eqnarray} 
	We assume the AdS/QCD scale parameter $\kappa=0.4$ GeV and an initial scale $\mu_{0}^2=0.32$ GeV$^2$. The parameters of the model are extracted using the electromagnetic properties of the proton, as discussed in detail in~\cite{Mondal:2017wbf}.

	\section{Extraction of GFFs}\label{GFFLFQDQ}
	The Form factors $A^{u+d}(Q^{2}),B^{u+d}(Q^{2})$ and $D^{u+d}(Q^{2})$ in the LFQDQ model can be parametrized in terms of structure integrals as~\cite{Chakrabarti:2020kdc,Chakrabarti:2015lba,Chakrabarti:2021mfd}
	\begin{eqnarray}
	A^{u+d}(Q^{2})=\mathcal{I}^{u+d}_{1}(Q^2), \hspace*{1cm} B^{u+d}(Q^{2})=\mathcal{I}^{u+d}_{2}(Q^2)
	\end{eqnarray}
	\begin{eqnarray}\label{DFF}
	D^{u+d}(Q^{2})=-\frac{1}{Q^{2}}\left[2M^{2}{\mathcal{I}}^{u+d}_{1}(Q^2)-Q^{2}{\mathcal{I}}^{u+d}_{2}(Q^2)-{\mathcal{I}}^{u+d}_{3}(Q^2)\right],
	\end{eqnarray}
where the full mathematical expressions of the integrals
 $\mathcal{I}_{i}^{u+d}(Q^{2})$ are given in~\cite{Chakrabarti:2020kdc,Choudhary:2022den}
It turns out that the form factor $D^{u+d}(Q^{2})$ can be parameterized by the multipole function as \cite{Chakrabarti:2021mfd},
	\begin{eqnarray}\label{Dfit}
D^{u+d}(Q^{2})=\frac{a}{(1+b  Q^{2})^{c}},
	\end{eqnarray}
where these evolved fitted parameters $a$, $b$ and $c$ are given as $a=D^{u+d}(0)=-1.521$, $b=0.531$ and  $c=3.026$~\cite{Choudhary:2022den} while at initial scale $a=D^{u+d}(0)=-18.8359 $, $b=2.2823$ and  $c=2.7951 $ . In order to perform the scale evolution, we employed the higher-order perturbative parton evolution toolkit (HOPPET)~\cite{Salam:2008qg} with the Dokshitzer-Gribov-Lipatov-Altarelli-Parisi (DGLAP) equations of QCD with NNLO.


	\section{EMT distributions}
	The form factors appearing in matrix elements of the EMT encode spatial densities via Fourier transforms. The two-dimensional light front energy, angular momentum, pressure and shear distributions are related to the GFFs by following relations respectively: 
	
	\begin{eqnarray}\label{Mass}
		\mathcal{E}^{(2D)}(x_{\perp})=P^{+}\tilde{A}(x_{\perp}), \hspace*{0.6cm} \rho_{J}^{(2D)}(x_{\perp})=-\frac{1}{2}x_{\perp}\frac{d}{dx}\tilde{J}(x_{\perp}) 
	\end{eqnarray}
	
\begin{figure} 
		\centering
		\includegraphics[scale=0.38]{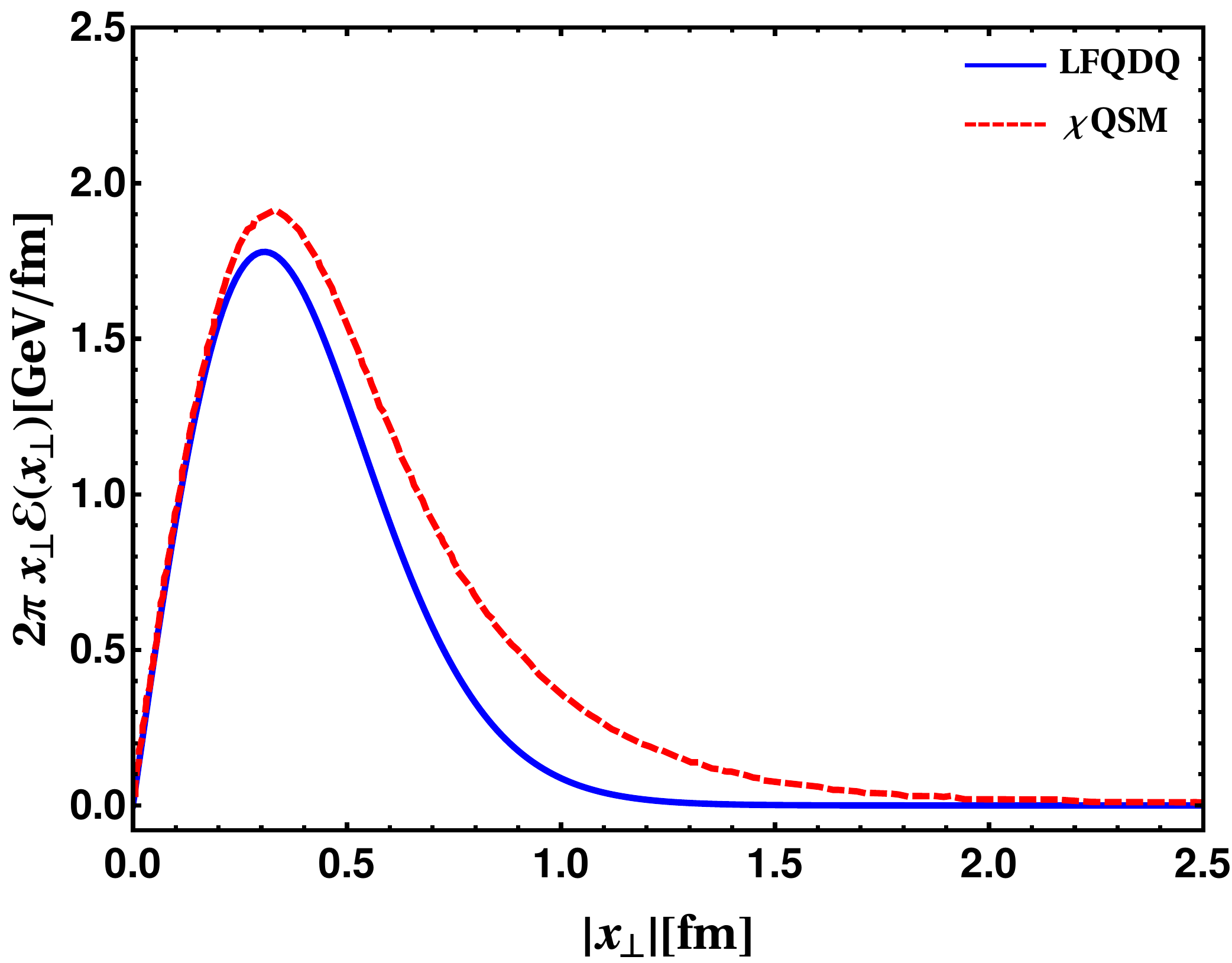} \hspace{0.5cm}
		\includegraphics[scale=0.38]{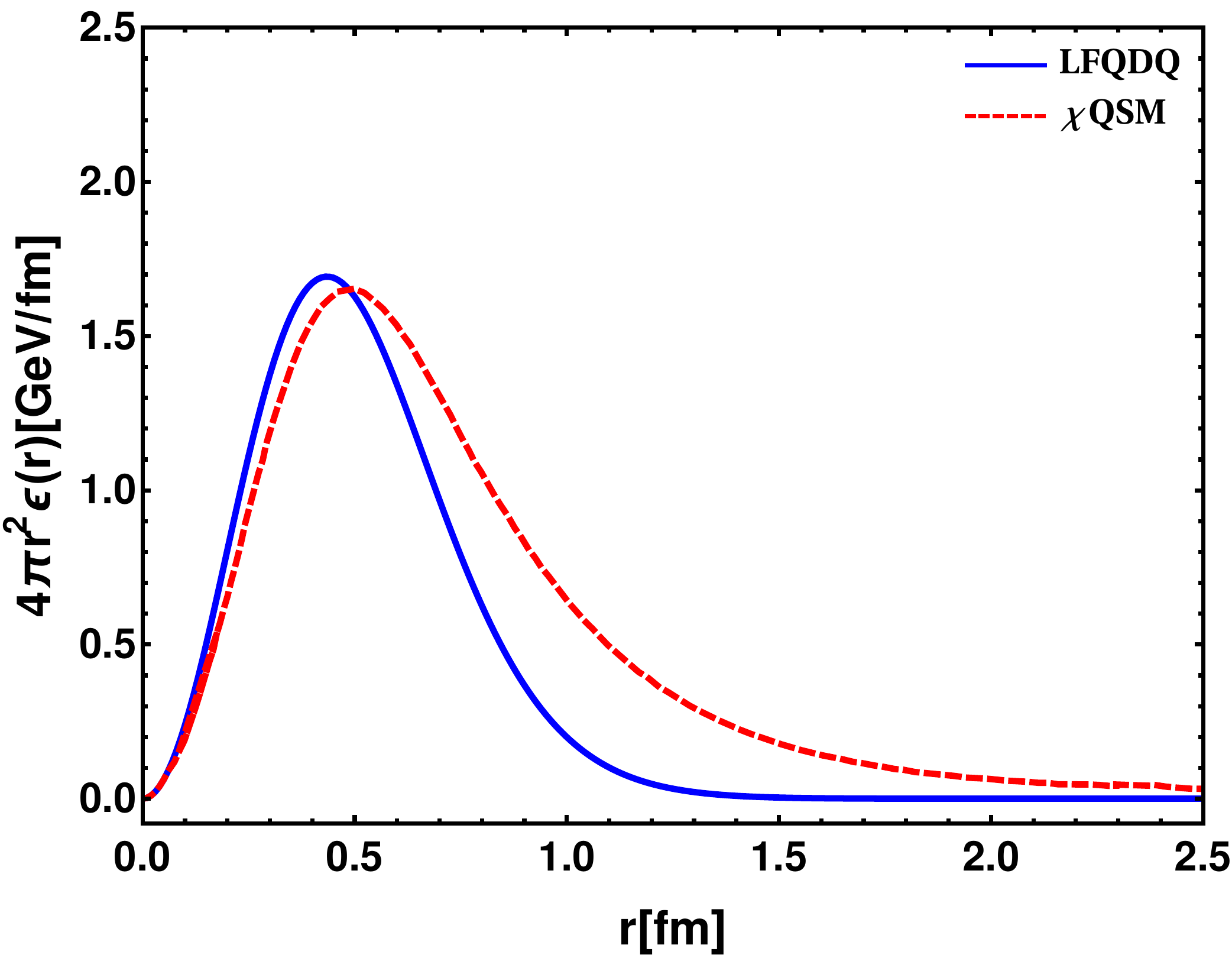}
		\caption{Left plot represents $ 2 \pi x_\perp $ weighted 2D mass distribution and right plot represents $ 4 \pi r^2 $ weighted 3D mass distribution at evolution scale $ \mu^2=4$ GeV$^2$. }
		\label{mass}
	\end{figure}
	 
	\begin{eqnarray} \label{2Dpressure and shear}
		p^{(2 D)}\left(x_{\perp}\right) &=\frac{1}{2 x_{\perp}} \frac{d}{d x_{\perp}}\left(x_{\perp} \frac{d}{d x_{\perp}} \tilde{D}\left(x_{\perp}\right)\right), \hspace*{0.6cm}	s^{(2 D)}\left(x_{\perp}\right) &=-x_{\perp} \frac{d}{d x_{\perp}}\left(\frac{1}{x_{\perp}} \frac{d}{d x_{\perp}} \tilde{D}\left(x_{\perp}\right)\right),
	\end{eqnarray}
	where 
	\begin{eqnarray}\label{2DFT}
		\tilde{F}(x_{\perp})=\int \frac{d^{2}\Delta}{(2\pi)^{2}}e^{-i\Delta_{\perp}.x_{\perp}}F(-\Delta_{\perp}^{2}),
	\end{eqnarray}

	\begin{figure}
	\centering
	\includegraphics[scale=0.38]{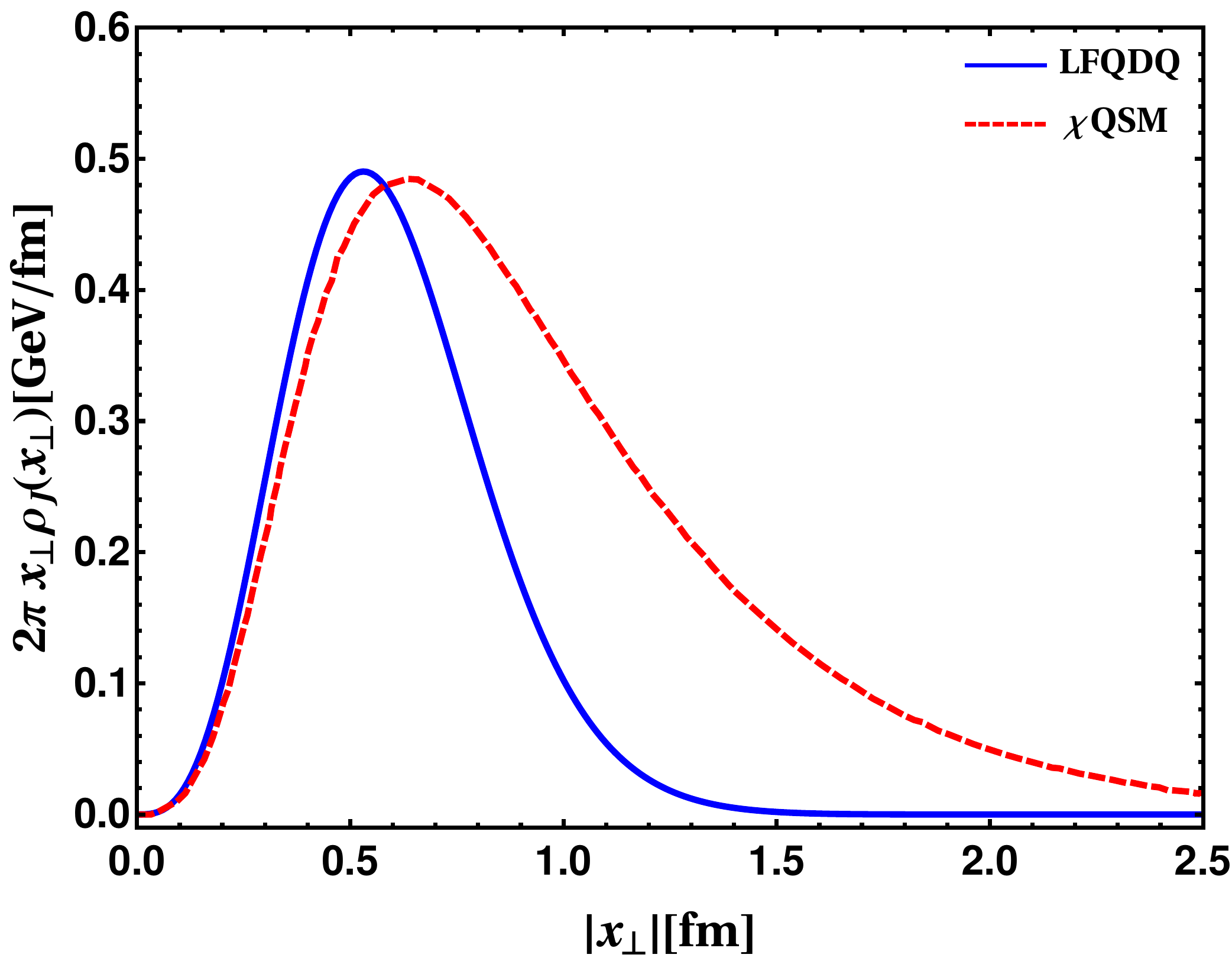}\hspace{0.5cm}
	\includegraphics[scale=0.38]{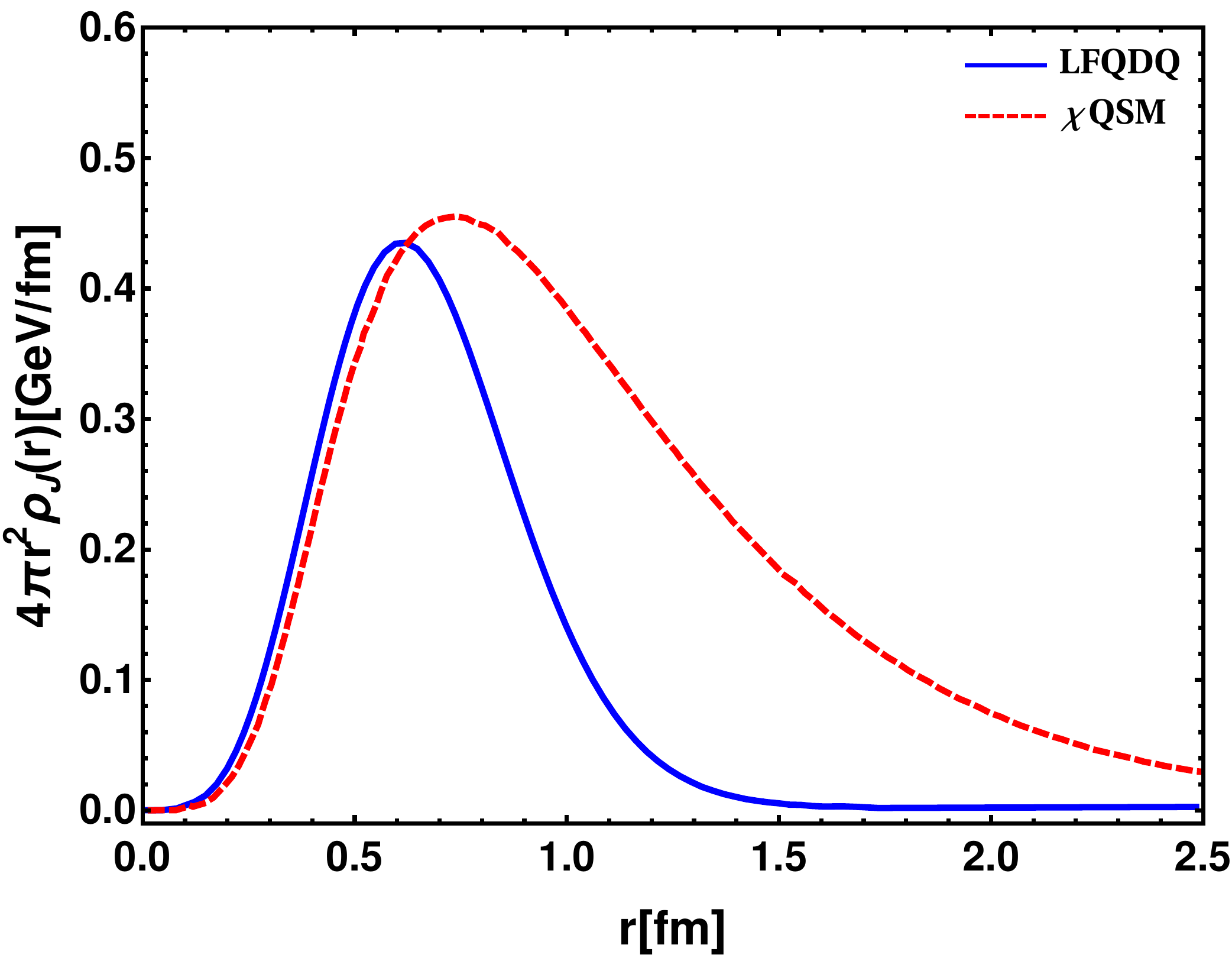}
	\caption{Left plot represents $ 2 \pi x_\perp $ weighted 2D angular momentum distribution and right plot represents $ 4 \pi r^2 $ weighted 3D angular momentum distribution at evolution scale $ \mu^2=4$ GeV$^2$.}
	\label{AGM}
\end{figure}
\begin{figure}
	\centering
	\includegraphics[scale=0.38]{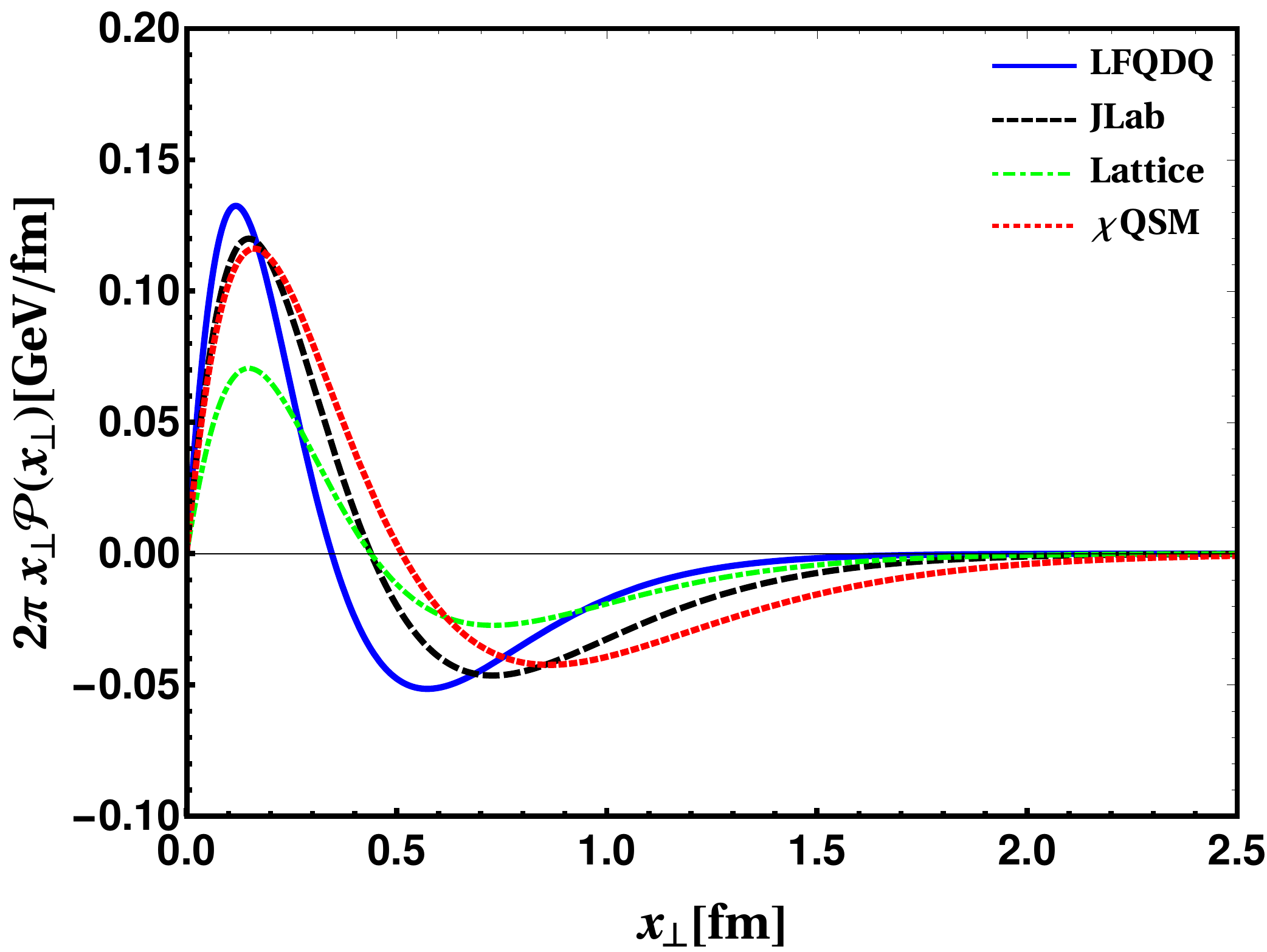}\hspace{0.5cm}
	\includegraphics[scale=0.38]{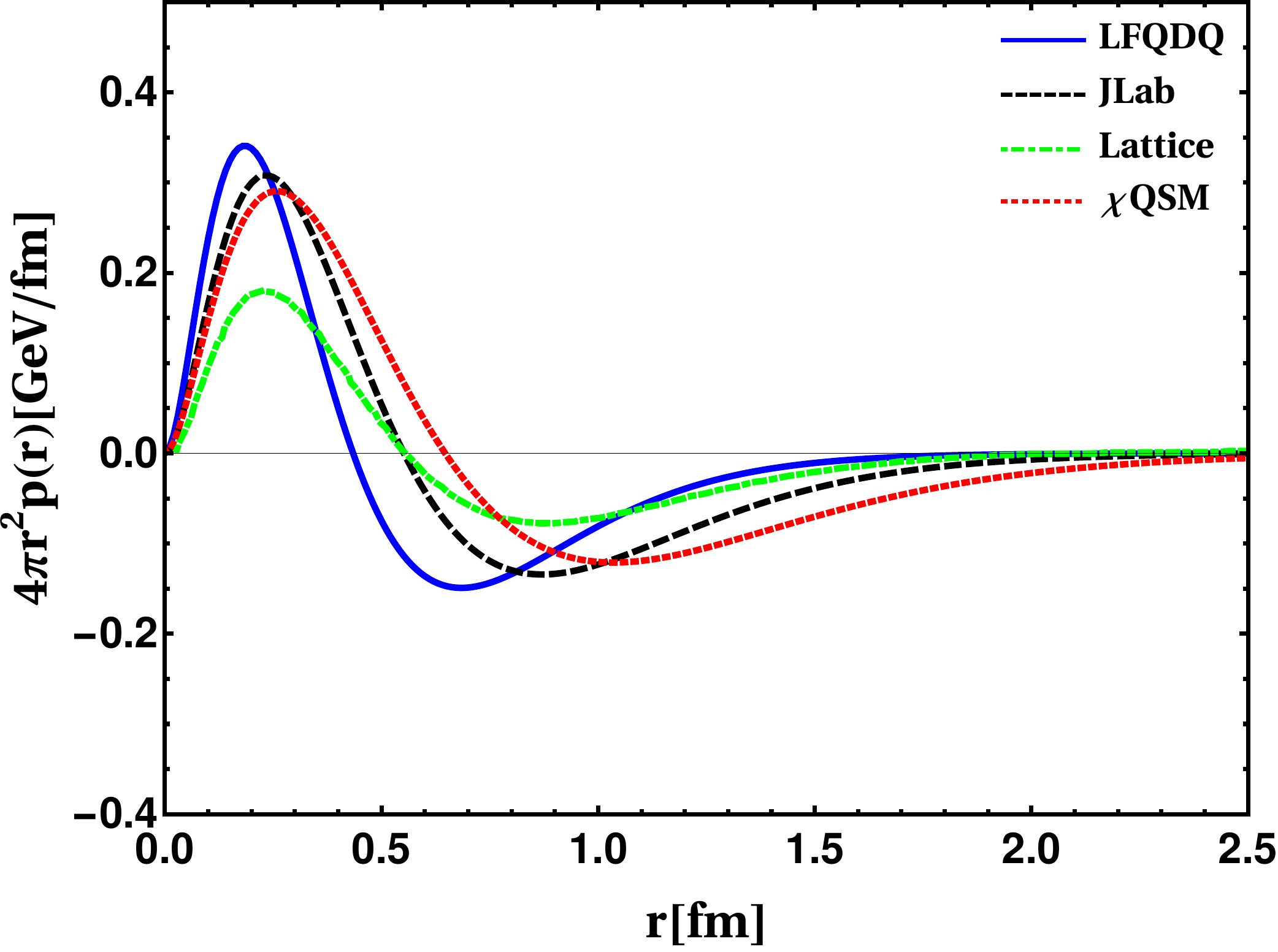}
	\caption{Left plot represents $ 2 \pi x_\perp $ weighted 2D pressure distribution and right plot represents $ 4 \pi r^2 $ weighted 3D pressure distribution at evolution scale $ \mu^2=4$ GeV$^2$. }
	\label{pressure}
\end{figure}
\begin{figure}
	\centering
	\includegraphics[scale=0.38]{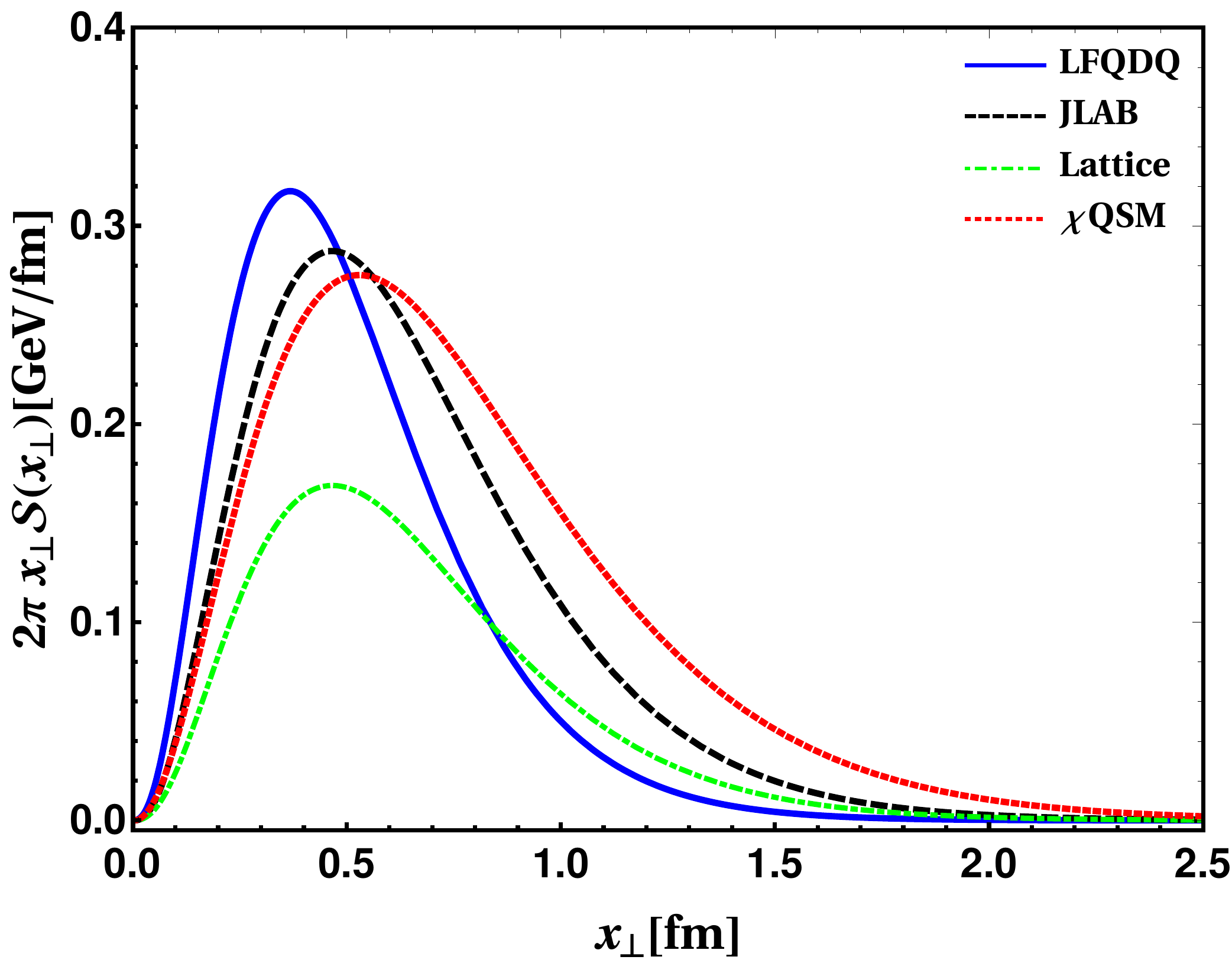} \hspace{0.5cm}
	\includegraphics[scale=0.38]{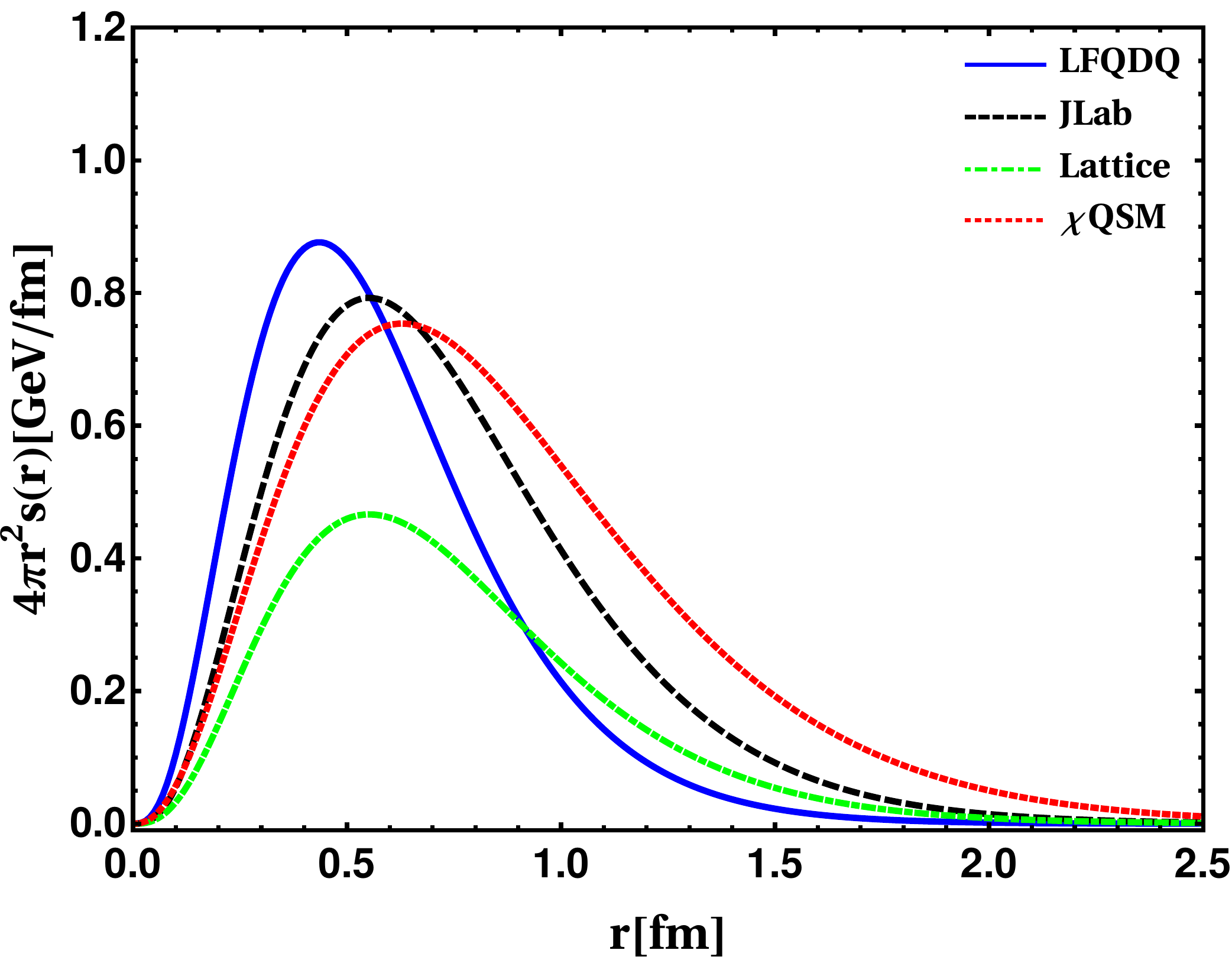}
	\caption{Left plot represents $ 2 \pi x_\perp $ weighted 2D shear force distribution and right plot represents $ 4 \pi r^2 $ weighted 3D shear force distribution at evolution scale $ \mu^2=4$ GeV$^2$.}
	\label{shear}
\end{figure}

	and $x_\perp$ and $\Delta_\perp$ are the position and momentum vectors in the two-dimensional plane that are perpendicular to the direction that the proton is travelling. Similarly, the three-dimensional EMT distributions in the Breit frame can be obtained by taking the GFFs and performing a three-dimensional inverse fourier transform on them. In this study~\cite{Choudhary:2022den}, we derived the 3D Breit frame EMT distributions from the 2D light front EMT distributions using the following inverse Abel transformation relations between them~\cite{Panteleeva:2021iip}. These distributions were derived from the 2D light front EMT distributions.
	\begin{eqnarray}\label{3Dmass}
		 \epsilon(r)&=-\frac{1}{\pi}\int_{r}^{\infty}\frac{dx_{\perp}}{x_{\perp}}\left(\mathcal{E}(x_{\perp})\right)\frac{1}{\sqrt{x_{\perp}^{2}-r^{2}}},
	\hspace*{0.6cm}	\rho_{J}(r)=-\frac{2}{\pi}r^{2}\int_{r}^{\infty}dx_{\perp}\frac{d}{dx_{\perp}}\left(\frac{\rho_{J}(x_{\perp})}{3x_{\perp}^{2}}\right)\frac{1}{\sqrt{x_{\perp}^{2}-r^{2}}}\nonumber \\
			s(r)& =-\frac{2}{\pi} r^{2} \int_{r}^{\infty} d x_{\perp} \frac{d}{d x_{\perp}}\left(\frac{\mathcal{S}\left(x_{\perp}\right)}{x_{\perp}^{2}}\right) \frac{1}{\sqrt{x_{\perp}^{2}-r^{2}}}, \hspace*{0.6cm}
			\frac{2}{3} s(r)+p(r) =\frac{4}{\pi} \int_{r}^{\infty} \frac{d x_{\perp}}{x_{\perp}} \mathcal{S}\left(x_{\perp}\right) \frac{1}{\sqrt{x_{\perp}^{2}-r^{2}}} 
	\end{eqnarray}
In Fig. \ref{mass} and Fig. \ref{AGM} we obtain the 2D and 3D mass and angular momentum distribution in our model and compared them with $\chi QSM$ model~\cite{Kim:2021jjf}, and showed that 2D and 3D distributions are Abel images of each other. Similarly in Fig.~\ref{pressure} and  Fig.~\ref{shear} our results for 2D and 3D pressure and shear distribution are presented and compared whih available results for $\chi QSM$~\cite{Kim:2021jjf}, JLab~\cite{Burkert:2018bqq,Burkert:2021ith} and Lattice simulations~\cite{Shanahan:2018nnv}, respectively. Our model results have higher peak values than the compared model results. Corresponding 2D and 3D radii for each distribution are shown in Table~\ref{table2}. Explicit calculation of each observable can be found in~\cite{Choudhary:2022den}. 
\begin{table}
		$$
	\begin{array}{lcccccc}
		\hline \hline \mathcal{P}(0)\left(\mathrm{GeV} / \mathrm{fm}^{2}\right) & \mathcal{E}(0)\left(\mathrm{GeV} / \mathrm{fm}^{2}\right) & \left(x_{\perp}\right)_{0}(\mathrm{fm}) & \left\langle x_{\perp}^{2}\right\rangle_{\text {J}}\left(\mathrm{fm}^{2}\right) & \left\langle x_{\perp}^{2}\right\rangle_{mass}\left(\mathrm{fm}^{2}\right) & \left\langle x_{\perp}^{2}\right\rangle_{\text {mech }}\left(\mathrm{fm}^{2}\right) \\
		\hline 0.354 & 1.54 & 0.34 & 0.38 & 0.21 & 0.167  \\
		\hline \hline p(0)\left(\mathrm{GeV} / \mathrm{fm}^{3}\right) & \varepsilon(0)\left(\mathrm{GeV} / \mathrm{fm}^{3}\right) & r_{0}(\mathrm{fm}) & \left\langle r^{2}\right\rangle_{\text {J }}\left(\mathrm{fm}^{2}\right) & \left\langle r^{2}\right\rangle_{mass}\left(\mathrm{fm}^{2}\right) & \left\langle r^{2}\right\rangle_{\text {mech }}\left(\mathrm{fm}^{2}\right)  \\
		\hline 4.76 & 2.02 & 0.43 & 0.51 & 0.32 & 0.251  \\
		\hline \hline
	\end{array}
	$$
	\caption{Different EMT distribution parameter values for the proton in 2D LF and 3D BF are as follows: ($\mathcal{E}(0)$,$\epsilon(0)$) -The energy distributions at the proton center, ($\mathcal{P}(0)$,$p(0)$) -pressure distribution at the proton center, ($(x_\perp)_0$,$r_0$)- nodal pressure points, and ($\langle x_\perp^2 \rangle$, $\langle r^2 \rangle$) the mean square radii of the mass, angular momentum, and mechanical.}
	\label{table2}
\end{table}

\begin{figure}
	\centering
	\includegraphics[scale=0.35]{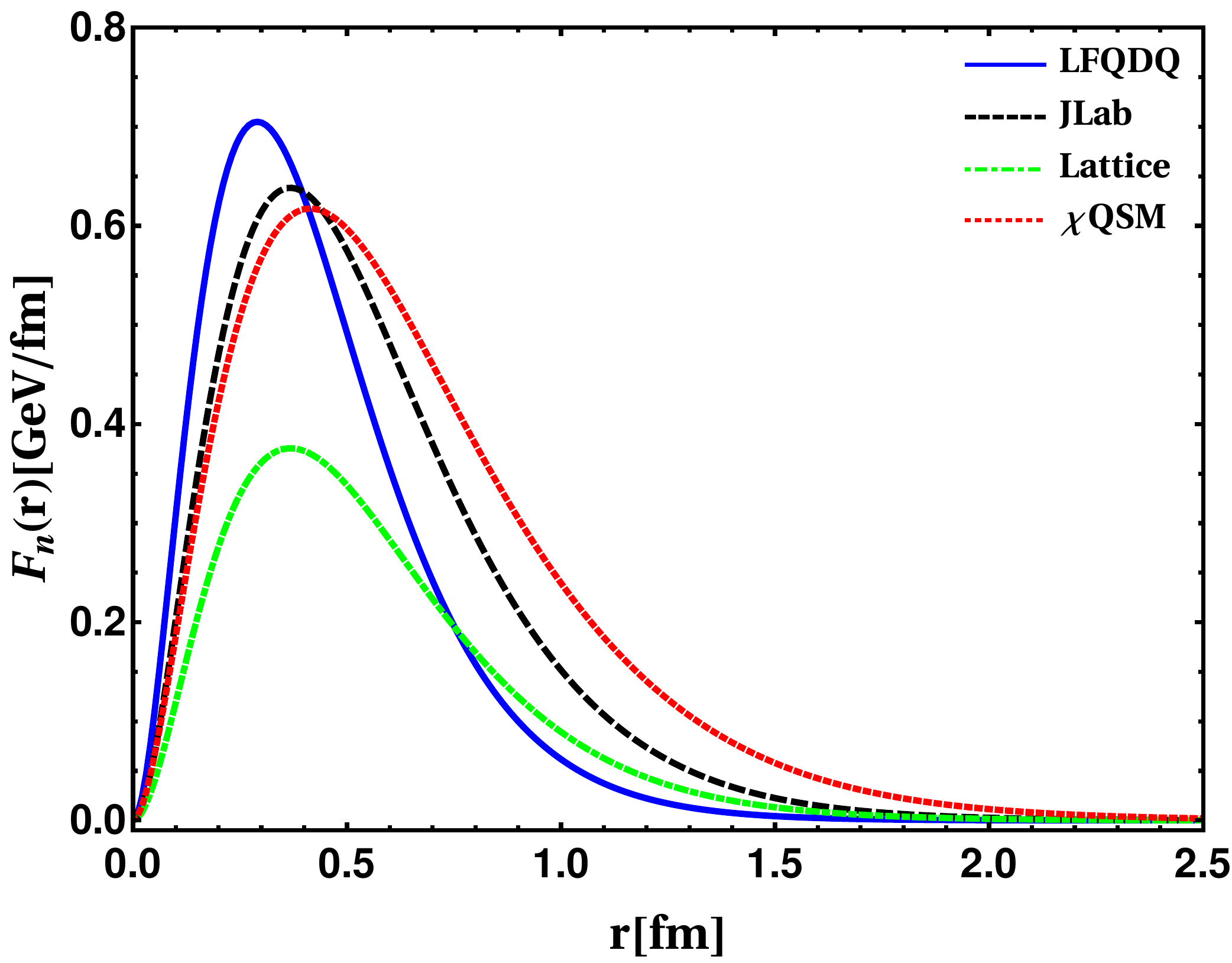} 
	\hspace{0.5cm}
	\includegraphics[scale=0.35]{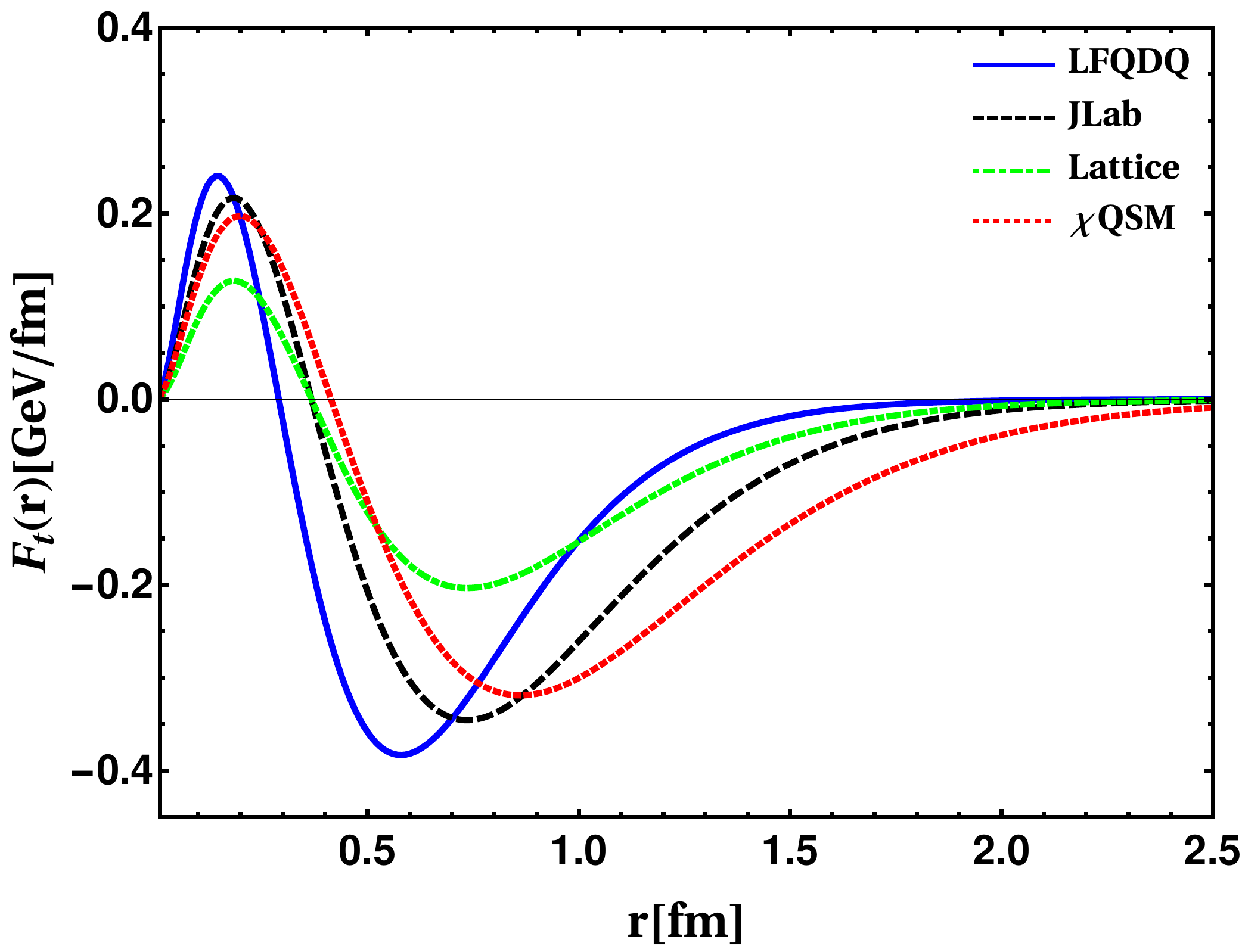}
	\caption{Three dimensional normal forces and tangential forces in the left and right panel respectively at evolution scale $ \mu^2=4$ GeV$^2$. }
	\label{Fn3DFt3D}
\end{figure}

The definitions of the tangential and normal force fields in three-dimensional Breit frame are given as~\cite{Polyakov:2018zvc,Kim:2021jjf},
\begin{eqnarray}\label{Fn3D-Ft3D}
	F_{n}(r)=4\pi r^{2}\left[\frac{2}{3}s(r)+p(r)\right], \hspace*{1cm} F_{t}(r)=4\pi r^{2}\left[-\frac{1}{3}s(r)+p(r)\right]
\end{eqnarray}  
In Fig.~\ref{Fn3DFt3D} we present our model results for normal and tangential force fields and compared them with other models. One nodal point in the tangential force shows the mechanical stability of the proton, whereas the positively distributed normal force field satisfies von Laue stability conditions. 
	\section*{Conclusion}\label{conclusion}	
	In this paper, the 2D LF distributions are evaluated in a scalar quark- diquark model of proton and then the 3D distributions are obtained in the model using the Abel transformation. Our results are compared with the $\chi QSM$, JLab and lattice predictions. The stability conditions are found to be satisfied with the LFQDQ model. The normal and shear force distributions are also evaluated in the LFQDQ model and are found to be consistent with lattice and other model predictions. 
	
	\section*{Acknowledgement} 
PC appreciates the opportunity to present this work at the "XXIX International Workshop on Deep-Inelastic Scattering and Related Subjects." AM acknowledges funding via a SERB-POWER Fellowship (file no. SPF/2021/000102) from the Science and Engineering Research Board (funded under Grant No. CRG/2019/000895).  
	
\bibliographystyle{unsrt}
	\bibliography{References1}
\end{document}